\documentclass[aps,prd,nopacs,floatfix,notitlepage,nofootinbib,twocolumn,a4paper]{revtex4}
\usepackage{amsfonts,amsmath,units,wasysym,epsfig,graphicx,verbatim,color,subfigure,graphicx,bm,mathrsfs,lipsum,hyperref,cleveref}
\usepackage{booktabs}
\usepackage[normalem]{ulem}  
\usepackage{multirow}

\begin{document}

\newcommand{\bhaskar}[1]{\textcolor{red}{ \bf BB: #1}}
\newcommand{\OKC}{The Oskar Klein Centre, Department of Astronomy, Stockholm University, AlbaNova, SE-10691 Stockholm,
Sweden}

\newcommand{\HU}{{Hamburger Sternwarte, Gojenbergsweg 112, D-21029 Hamburg, Germany}}

\title{PSR J0952-0607: Probing the Stiffest Equations of State and r-Mode Suppression Mechanisms}

\author{Zeyue Wu$^{\rm 1}$, Bhaskar Biswas$^{\rm 1}$, Stephan Rosswog$^{\rm 1, \rm 2}$}
\affiliation{$^{\rm 1}$\HU, $^{\rm 2}$\OKC}
 
\begin{abstract}
We analyze PSR J0952-0607, the most massive and fastest spinning neutron star observed to date, to refine constraints on the neutron star equation of state (EoS) and we investigate its robustness against r-mode instabilities. With a mass of \( 2.35 \pm 0.17 \, M_{\odot} \) and a spin frequency of 709.2 Hz, PSR J0952-0607 provides a unique opportunity to examine the effects of rapid rotation on the structure of a neutron star. Using a Bayesian framework, we incorporate the rotationally corrected mass of PSR J0952-0607, alongside PSR J0740+6620’s static mass measurement, to constrain the EoS. Our findings demonstrate that neglecting rotational effects leads to biases in the inferred EoS, while including the neutron star spin produces tighter constraints on pressure-density and mass-radius relations. Additionally, we explore the r-mode instability window for PSR J0952-0607 under the assumption of both rigid and elastic crust models  and find that a rigid crust allows a higher stable temperature range, whereas an elastic crust places the star within the instability window under certain thermal insulation conditions.



\end{abstract}

\maketitle

\section{Introduction}

The recent discovery of PSR J0952-0607~\cite{Bassa:2017zpe,Romani:2022jhd}, the most massive and fastest-spinning neutron star (NS) known, provides a unique opportunity to probe the physics of ultra-dense matter. With a mass of \(2.35 \pm 0.17\) solar masses and a spin frequency of 709.2 Hz (period 1.41 ms), PSR J0952-0607 challenges current models of the neutron star equation of state (EoS)~\cite{Lattimer_2016,Oertel_2017,Baym_2018}. The relationship between pressure and density at supranuclear densities remains uncertain due to complexities in strong nuclear interactions, and observations of massive, fast-spinning neutron stars like PSR J0952-0607 provide a valuable opportunity to refine these models. 

The Chandrasekhar-Friedman-Schutz (CFS) instability~\citep{Chandrasekhar:1992pr,Friedman:1978hf} occurs when a retrograde mode in the rotating frame becomes prograde in the inertial frame, which leads to the exponential growth of gravitational radiation. This instability affects various modes, including $f$-modes and $r$-modes. In relativistic stars, the  $f$-mode instability can already emerge at  $m=2$ , in contrast to earlier Newtonian results, which requires  $m = 3$  or $m = 4 $. This effect is particularly pronounced in massive neutron stars~\citep{Stergioulas:1997ja,Morsink_1999}. The $r$-mode instability, in particular, is inherently tied to the CFS mechanism and is generally unstable at any spin frequency~\citep{Lindblom:1998wf,1999ApJ...510..846A}. Its growth depends on the competition between the r-mode growth timescale and the damping effects of various mechanisms~\cite[see, e.g.,][]{Madsen:1999ci,2000PhRvD..61j4003L,Sa:2003jw,Sa:2004gn,10.1111/j.1365-2966.2009.14963.x,Haskell_2010,2011PhRvL.107j1102G,Wen:2011xz,Vidana:2012ex,Haskell:2012vg,Moustakidis:2012fn,Papazoglou:2015zqa,Ofengeim:2019fjy}.

 The secondary component of the gravitational wave event GW190814~\cite{Abbott:2020khf} presents a fascinating case for studying extreme compact objects. Although the spin of the secondary component is unconstrained, its nature remains debated: it could be either a superfast pulsar or a low-mass black hole. Biswas et al.~\cite{Biswas:2020xna} have highlighted that for the secondary to be a superfast pulsar, its rotational instabilities—such as r-mode instabilities—would need to be suppressed. Without such suppression, gravitational radiation due to these instabilities would quickly lead to the exponential loss of rotational energy and would therefore favor a black hole interpretation. However, the absence of a precise spin measurement for this object limits the ability to test r-mode instability mechanisms in this system.

In contrast, PSR J0952-0607, with its precisely known spin, offers a unique opportunity to study the mechanisms of r-mode instability in a rapidly rotating neutron star. The precise spin measurement, combined with its extreme mass, allows us to directly probe the stability of high-spin neutron stars and assess the role of EoS properties and crustal characteristics in r-mode suppression. This makes PSR J0952-0607 an ideal laboratory for exploring the interplay between rapid rotation, dense matter physics, and gravitational-wave emission.

In this study, we use the mass and rapid rotation of PSR J0952-0607 to place stringent constraints on the neutron star EoS using Bayesian inference. Our approach incorporates the star’s rapid spin into the EoS inference process, as high rotation rates can increase a neutron star’s maximum mass substantially \cite{1994ApJ...424..823C}. By including the spin measurement, we aim to improve the reliability of EoS constraints. We will also demonstrate how neglecting the spin measurement can skew results and mischaracterize the EoS.

The r-mode instability window—defined in the spin frequency versus temperature plane~\cite{Andersson:2000mf}—depends sensitively on the EoS of neutron-rich matter~\cite{PhysRevC.85.025801,Papazoglou:2015zqa}. For this purpose, we use all possible equations of state inferred from our Bayesian analysis that satisfy the mass measurement of PSR J0952-0607 to compute the r-mode instability window. Another  key ingredient is the physical state of the neutron star crust: whether it is rigid or elastic can drastically affect the r-mode stability properties. 
These dependencies are discussed in detail in the reviews of
 Andersson and Kokkotas~\cite{Andersson:2000mf}, Haskell~\cite{Haskell:2015iia},  and Kokkotas and Schwenzer~\cite{Kokkotas:2015gea}. We examine here the crucial role that the physical state of the crust plays in determining the extent of viscous damping, which directly affects the boundaries of the r-mode instability window.

The rest of this paper is organized as follows. In Section~\ref{sec:review}, we review the modelling ingredients that enter our analysis, including the piecewise polytropic EoS model, the mass-radius relationships for both non-rotating and rotating neutron stars, and the theoretical framework for the r-mode instability. Section~\ref{sec:bayes-method} presents our Bayesian inference methodology, detailing the likelihood functions, prior distributions, and sampling techniques, along with a discussion of the scope and limitations of our approach. Section~\ref{sec:result} presents our results, including constraints on neutron star EoS parameters, mass-radius relationships, and an analysis of the r-mode instability boundaries for PSR J0952-0607 under different assumptions for the crust. Finally, Section~\ref{sec:conclusion} concludes the paper with a summary of our findings and their implications for neutron star structure, rotational stability, and future observational constraints on the EoS.

\label{introduction}

\section{Modelling ingredients}
\label{sec:review}
In this work, 
we adopt geometrized units ($G = c = 1$) in Subsections \ref{subsec:PPEoS} and \ref{subsec:TOV}. However, in Subsections \ref{subsec:RotatingNs} and \ref{subsec:rmode}, due to their closer theoretical connection to the physical properties of neutron stars, we follow the convention of using CGS units.
\subsection{Piecewise polytropic EoS}
\label{subsec:PPEoS}
In this work, we employ a piecewise polytropic (PP) EoS parameterization~\cite{Read:2008iy} to model the dense matter inside neutron stars. In the PP model, the EoS is divided into several density segments, each described by its own polytropic relation. This allows for a  flexible and accurate representation of the NS EOS across different density ranges. For the rest-mass density, $\rho$, in the interval between $[\rho_i, \rho_{i+1}]$, the pressure, $p$, and energy density, $\epsilon$, are given by:
\begin{eqnarray}\label{eq:poly_eos}
    p(\rho) &=& K_i\rho^{\Gamma_i} \\
    \epsilon(\rho) &=& \frac{K_i}{\Gamma_i+1}\rho^{\Gamma_i} +(1+a_i)\rho \, ,
\end{eqnarray}
where the parameters $K_i$ and $a_i$ are determined by enforcing continuity of the energy density and pressure across the dividing densities:
\begin{eqnarray}
    K_{i+1}&=& K_i\rho^{\Gamma_i-\Gamma_{i+1}} \label{eq:Ki}\\
    a_{i} &=& \frac{\epsilon(\rho_{i-1})}{\rho_{i-1}} - 1  - \frac{K_{i}}{\Gamma_{i} - 1}
\rho_{i-1}^{\Gamma_{i} - 1}.  \label{eq:ai}
\end{eqnarray}
Our PP model contains four parameters $\log p_1, \Gamma_1, \Gamma_2, \Gamma_3$, where $p_1$ is the pressure at the first dividing density, and the $\Gamma_i$ denote the polytropic indices separated by two fixed transition densities at $1.8 \rho_0$ and $3.6 \rho_0$ respectively, where $\rho_0=2.75 \times 10^{14} \, {\rm g/cm^3}$ is the nuclear saturation density.

\subsection{Mass and Radius of nonrotating Neuton Stars}
\label{subsec:TOV}
The structure of neutron stars is found by solving the Tolman-Oppenheimer-Volkoff (TOV) equations~\cite{PhysRev.55.364, PhysRev.55.374}, which are derived from the general relativistic  conditions for hydrostatic equilibrium. With $G = 1$ and $c = 1$, the TOV equations simplify to:
\begin{equation}
\frac{dp(r)}{dr} = -\frac{(\epsilon(r) + p(r))(m(r) + 4 \pi r^3 p(r))}{r (r - 2 m(r))}
\end{equation}
\begin{equation}
\frac{dm(r)}{dr} = 4 \pi r^2 \epsilon(r)
\end{equation}
where  $p(r)$ is the pressure given by the used EOS, $\epsilon(r)$ is the energy density, and $m(r)$ is the mass enclosed within the radius $r$. 
The radius $R$ of the neutron star is determined by the condition that the pressure vanishes, i.e., $p(R) = 0$.

\subsection{Mass and Radius of Rotating Neutron Stars}
\label{subsec:RotatingNs}
%

\begin{figure}[h!]
\centering
\includegraphics[width=0.9\linewidth]{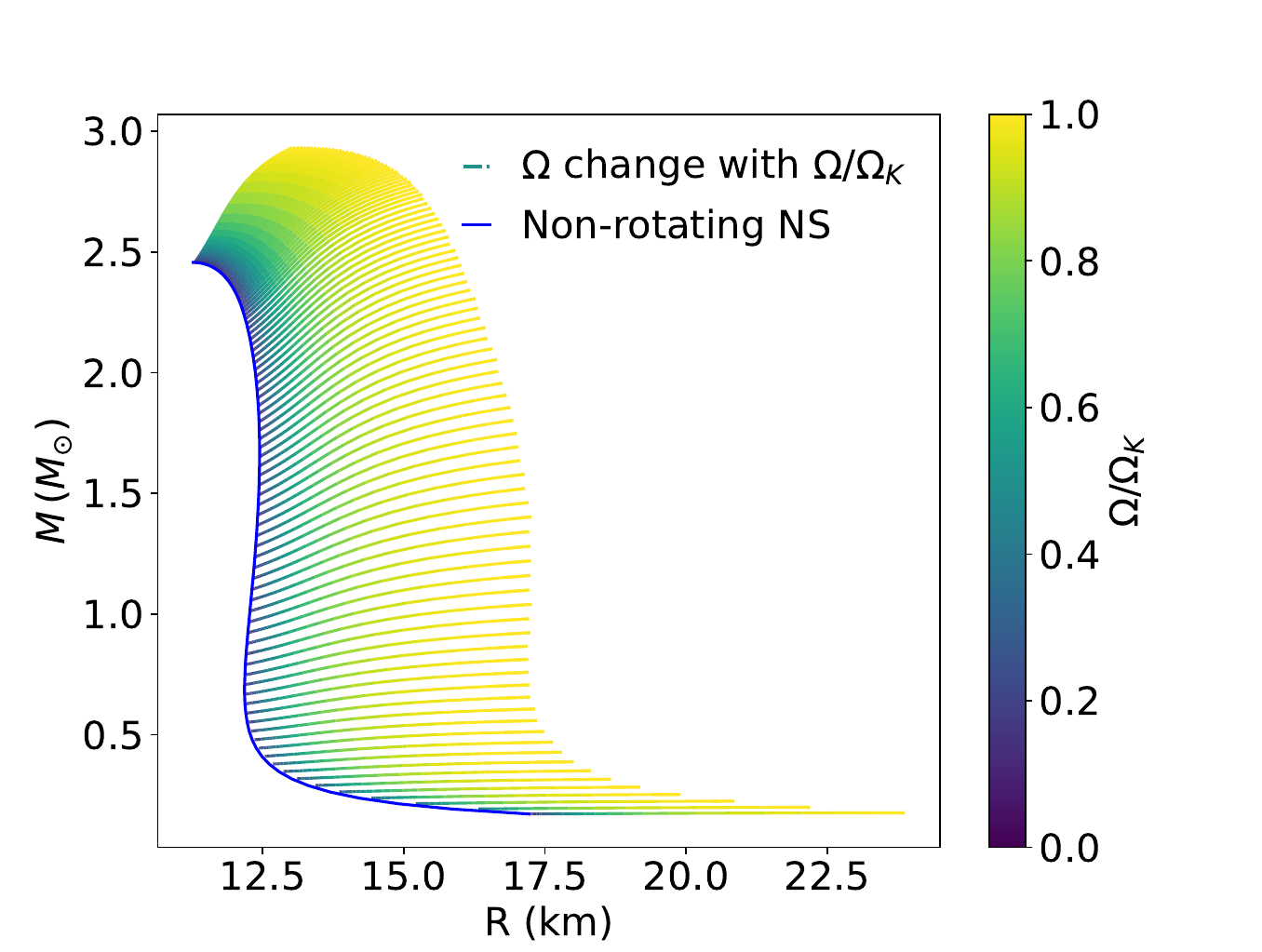}
\caption{Impact of the NS spin frequency on the mass-radius sequences of NSs for the example of the MPA1 EoS.}
\label{TOV_Omega}
\end{figure}

The effects of rotation can be approximated via polynomials
in the compactness \( C_* = \frac{M}{R} \)\cite{Konstantinou:2022vkr}, where \( M \) and \( R \) are the mass and radius of a non-rotating neutron star. Denoting the mass and equatorial radius when the neutron star is rotating as  $M_e$  and  $R_{e}$, one finds \cite{Konstantinou:2022vkr}
\begin{equation} 
\Omega_k = \Omega_* \sum_{i=0}^4 a_i C_*^i,
\label{omega_k}
\end{equation}
\begin{equation}
\begin{aligned}
\frac{R_{e}}{R} &= 1 + \left(e^{A_r \Omega_n^2} - 1 + B_r \left[\ln\left(1 - \left(\frac{\Omega_n}{1.1}\right)^4\right)\right]^2\right) \times \\
&\left(1 + \sum_{i=1}^5 a_{r,i} C_*^i\right), \\
\frac{M_{e}}{M} &= 1 + \left(e^{A_m \Omega_n^2} - 1\right) \times \left(1 + \sum_{i=0}^4 a_{m,i} C_*^i\right),
\label{M_R_rotating}
\end{aligned}
\end{equation}
where \(\ \Omega_k \) is the limiting (``Kepler") angular frequency,
\(\Omega_n = \frac{\Omega}{\Omega_k}\) is the normalized angular velocity, with \(\Omega\) being the angular velocity of the rotating neutron star and $\Omega_* = \frac{GM}{R^3}$. The coefficients \(a_i\) and other parameters can be found in \cite{Konstantinou:2022vkr}. 

To illustrate the impact of the neutron star spin on the 
mass-radius relation, we use the MPA1~\cite{MPA1} EoS as an example. We assume that the neutron star's maximum rotation rate corresponds to its Keplerian angular velocity \(\Omega_k\) as given in equation (\ref{omega_k}). The resulting mass-radius curves are shown in Figure \ref{TOV_Omega}.

\subsection{The r-mode instability}
\label{subsec:rmode}
The r-mode instability is a phenomenon dominated by the Coriolis force, which drives oscillations in a rotating neutron star, leading to gravitational wave emission and a gradual decrease in the star's rotation rate. The velocity perturbation of a rotating Newtonian star induced by r-mode is given by:
\begin{equation}
\delta \vec{v}=\alpha R\Omega \left(\frac{r}{R}\right)^{l}\vec{Y}^{B}_{lm}e^{i\omega t-\frac{t}{\tau}}
\label{delta_v}
\end{equation}
where \( R \) and \( \Omega \) are the radius and angular velocity of the unperturbed neutron star, \(\tau\) is the time scale, and \(\vec{Y}^{B}_{lm}\) is the magnetic type vector spherical \(lm\) harmonic \cite{Owen:1998xg}. The quantity \(\alpha\) is an arbitrary constant, serving as a factor to adjust the amplitude. In this work, we specifically select the fundamental r-mode, defined by $l=m=2$, as the focus of our subsequent analysis. The energy associated with the r-mode oscillation can be written as \cite{Owen:1998xg}:
\begin{equation}
\widetilde{E}=\frac{1}{2}\int \left[\rho\delta \vec{v}\cdot\delta \vec{v}^{*}+ \left(\frac{\delta \rho }{\rho}-\delta \Phi \right)\delta \rho^{*}\right]d^3 r
\label{waveE}
\end{equation}
where \(\rho\) is the density of the neutron star and \(\delta \rho\) represents its perturbation and  $\delta\vec{v}^{*}$ and $\delta \rho^{*}$ represent the complex conjugates of $\delta\vec{v}$ and $\delta \rho$.  \(\delta \Psi\) is a quantity which is proportional to the Newtonian  gravitational potential perturbation \(\delta \Phi\). It is the solution~\cite{Owen:1998xg} to the ordinary differential equation:
\begin{equation}
\begin{aligned}
\frac{d^2 \delta \Psi(r)}{dr^2} + \frac{2}{r} \frac{d \delta \Psi(r)}{dr} + 
\left[ 4\pi G \rho \frac{d\rho}{dp} - \frac{(l+1)(l+2)}{r^2} \right] \delta \Psi(r) \\
= - \frac{8 \pi G l}{2l+1} \sqrt{\frac{l}{l+1}} \rho \frac{d\rho}{dp} 
\left( \frac{r}{R} \right)^{l+1}.
\end{aligned}
\label{eq:delta_Psi}
\end{equation} 
Since the perturbations vary with 
$\exp (i\omega t-\frac{t}{\tau})$
the energy of the r-mode satisfies the relation \cite{Owen:1998xg}:
\begin{equation}
\frac{d\widetilde{E}}{dt}=-\frac{2\widetilde{E}}{\tau}
\label{waveE-new}
\end{equation}

Using Equation (\ref{waveE-new}), we can derive the dependency of a neutron star's spin $\Omega$ on temperature $T$ as:
\begin{equation}
\frac{d\Omega}{dT} = \frac{dt}{dT} \frac{\Omega}{\tau},
\label{Omega}
\end{equation}
where $\frac{dt}{dT}$ can be expressed through various cooling models, and the characteristic timescale \(\tau\) is also derived using Equation (\ref{waveE-new}). In particular, the damping timescales associated with gravitational radiation, shear, and bulk viscosity effects influence this decay rate.

The characteristic gravitational radiation timescale \(\tau_{gr}\)  can be expressed as \cite{Owen:1998xg}:
\begin{equation}
\begin{split}
\frac{1}{\tau_{gr}} = & -\frac{32\pi G \Omega^{2l+2}}{c^{2l+3}} \cdot 
\frac{{(l-1)}^{2l}}{\left[(2l+1)!!\right]^2} \\
& \times \left( \frac{l+2}{l+1} \right)^{2l+2} 
\int_{0}^{R} \rho r^{2l+2} dr.
\end{split}
\label{time scale gr}
\end{equation}
Here $(2l+1)!!$ is the double factorial of $2l+1$. Timescale \(\tau_s\) corresponds to shear viscosity, which depends significantly on whether the neutron star has a rigid or  an elastic crust. For a neutron star with a rigid crust, r-mode oscillations are damped at the boundary between the inner crust and the outer core, leading to the following shear timescale:
\begin{equation}
\begin{split}
\tau_{\rm s,rigid} = & \frac{1}{2\Omega} 
\frac{2^{l+\frac{3}{2}}(l+1)!}{l(2l+1)!!C_l} \\
& \times \left(\frac{2\Omega R_0^2 \rho_0}{\eta_{nn}(\rho_0)}\right)^{1/2} \\
& \times \int_{0}^{R_0} \frac{\rho}{\rho_0} 
\left(\frac{r}{R_0}\right)^{2l+2} \frac{dr}{R_0},
\end{split}
\label{time scale s_rigid}
\end{equation}
where \(C_l\) is a coefficient (\(C_2=0.80411\)), and \(\rho_0\) is the neutron saturation density, which we consider as the boundary density as the boundary density between the neutron star's inner crust and outer core, and \(R_0\) is the radius of the neutron star's outer core \cite{Zhou:2020xan}. 

 For an elastic crust, in turn, the crust participates fully in the r-mode oscillations and its shear timescale is given by \cite{Zhou:2020xan}:
\begin{equation}
\begin{split}
\frac{1}{\tau_{\rm s,elastic}} = & (l-1)(2l+1) \\
& \times \int_{0}^{R} \eta_{\rm elastic} r^{2l} dr \\
& \times \left( \int_{0}^{R} \rho r^{2l+2} dr \right)^{-1}.
\end{split}
\label{time scale s}
\end{equation}

The shear viscosity \(\eta\) varies with location in the neutron star. For a rigid crust, it is dominated by neutron viscosity, given by \cite{Zhou:2020xan}:
\begin{equation}
\eta_{nn}=347\rho^{\frac{9}{4}}T^{-2}.
\label{eta}
\end{equation}
In an elastic crust, shear viscosity near the outer crust is due to electron viscosity  \cite{Zhou:2020xan}:
\begin{equation}
\eta_{ee}=6 \times 10^6 \rho^2 T^{-2}.
\label{eta_ee}
\end{equation}
As mentioned in \ref{sec:review}, both $\eta_{nn}$ and $\eta_{ee}$ are expressed in CGS units.  Closer to the inner crust, neutron viscosity prevails. The combined shear viscosity in an elastic crust is assumed to vary linearly between these regions:
\begin{equation} 
\eta_{elastic} = \frac{\rho_0 - \rho}{\rho_0 - \rho_{out}}\eta_{ee} + \frac{\rho - \rho_{out}}{\rho_0 - \rho_{out}}\eta_{nn},
\label{eta_elastic}
\end{equation} 
where \(\rho_{out} = 2.62 \times10^{12} \, \text{g} \, \text{cm}^{-3}\) is the density at the outer crust-inner crust boundary, defined by the SLy~\cite{Douchin:2001sv} EoS used in the PP EoS parameterization. 

The bulk viscosity term \(\tau_b\) is calculated as:
\begin{equation}
\frac{1}{\tau_{b}} \approx \frac{4R^{2l-2}}{(l+1)^2} \int \xi \left|\frac{\delta \rho}{\rho}\right|^2 d^3 x \left(\int_{0}^{R}\rho r^{2l+2}dr\right)^{-1},
\label{time scale b}
\end{equation}
where \(\xi\) is the bulk viscosity of hot neutron star matter, expressed in cgs units as \cite{Owen:1998xg}:
\begin{equation}
\xi=6 \times 10^{-59} \left( \frac{l+1}{2\Omega} \right)^2 \rho^2 T^6,
\label{xi}
\end{equation}
and the density perturbation \(\delta \rho\) is given by:
\begin{equation}
\begin{split}
\delta \rho = & \alpha R^2 \Omega^2 \rho \frac{d\rho}{dp} \\
& \times \left[ \frac{2l}{l+1} \sqrt{\frac{l}{l+1}} 
\left( \frac{r}{R} \right)^{l+1} + \delta \Psi(r) \right] \\
& \times Y_{l+1} e^{i\omega t},
\end{split}
\label{deltatrho}
\end{equation}

The total timescale \(\tau\) is the combined effect of each damping mechanism which depends on neutron star’s spin $\Omega$ and temperature $T$,  can be evaluated analytically:
\begin{equation}
\begin{split}
\frac{1}{\tau(\Omega, T)} = & \frac{1}{\widetilde{\tau}_{gr}} 
\left( \frac{\Omega^2}{\pi G \bar{\rho}} \right)^{l+1} \\
& + \frac{1}{\widetilde{\tau}_{s}} 
\left( \frac{10^9 \, \text{K}}{T} \right)^2 \\
& + \frac{1}{\widetilde{\tau}_{b}} 
\left( \frac{T}{10^9 \, \text{K}} \right)^6 
\left( \frac{\Omega^2}{\pi G \bar{\rho}} \right),
\end{split}
\label{time scale new}
\end{equation}
 where \(\bar{\rho}\) is the average density of the neutron star  \(\bar{\rho} = M / \frac{4}{3} \pi R^3\) and \(\widetilde{\tau}\) represents fiducial timescales evaluated at $\Omega$  and at $T$. This analytical evaluation of  \(\tau(\Omega, T)\)  further enables calculate the r-mode instability region.

The exponential time dependence of the r-mode energy, given by \(e^{i\omega t - \frac{t}{\tau}}\), allows us to determine whether r-modes will grow or decay. When \(\tau\) is positive, r-mode energy \(\widetilde{E}\) decays exponentially over time, stabilizing the angular velocity. However, when \(\tau\) is negative, \(\widetilde{E}\) grows, driving r-mode instability and reducing angular velocity due to gravitational wave emission \cite{Owen:1998xg}. This framework enables us to delineate the r-mode instability region in terms of the temperature (\(T\)) and angular velocity (\(\Omega\)) of a neutron star. Solving for the boundary, \(\tau(\Omega, T) = 0\), provides a curve of critical temperatures and angular velocities, \(T_c\) and \(\Omega_c\), which define the r-mode instability window.\\

\section{Bayesian Inference Methodology}
\label{sec:bayes-method}
Our focus here is on the impact that the mass and spin measurements of PSR J0952-0607 have on the inferred neutron star EoS. PSR J0952-0607, a rapidly rotating neutron star with a spin frequency of 709.2 Hz, measured using the LOFAR telescope and phase folding techniques as described in \cite{Bassa:2017zpe}, with measurement errors on the order of nanoseconds and thus negligible, requires us to account for rotational effects when inferring its mass. To assess how this impacts the inferred EoS, we compare the results from PSR J0952-0607 with those of PSR J0740+6620, a neutron star with a mass of \( 2.08 \pm 0.07 \, M_\odot\) and a spin frequency of 340 Hz~\cite{Cromartie:2019kug,Fonseca:2021wxt}, providing a high-mass benchmark. 

Our goal is to infer the posterior distribution of EoS parameters by combining these two neutron stars' mass and spin measurements in a Bayesian framework. We will evaluate how these measurements improve the EoS inference without considering constraints from gravitational waves~\cite{TheLIGOScientific:2017qsa,Abbott:2018exr,Abbott:2020uma} or mass-radius measurements from NICER~\cite{Riley:2019yda,Miller:2019cac,Riley:2021pdl,Miller:2021qha,Choudhury:2024xbk}, focusing purely on the impact of these pulsars.

\subsection{Likelihood and Priors}

The combined likelihood for both PSR J0952-0607 and PSR J0740+6620 is modeled as a product of the individual likelihoods for the two pulsars. Each likelihood is modeled as a Gaussian distribution based on the observed mass, spin (for PSR J0952-0607), and the EoS parameters.
The total likelihood is:

\begin{equation}
    \mathcal{L}(\text{data} | \theta) = \mathcal{L}_{0952}(\text{data} | \theta) \cdot \mathcal{L}_{0740}(\text{data} | \theta),
\end{equation}

where \( \theta \) represents the EoS parameters.

For PSR J0740+6620, we assume a simple mass likelihood without rotational corrections:

\begin{equation}
    \mathcal{L}_{0740} \propto \exp\left(-\frac{(M_{\text{0740}}(\theta) - M_{\text{obs,0740}})^2}{2\sigma_{0740}^2}\right),
\end{equation}

where \( M_{\text{0740}}(\theta) \) is the mass predicted from the EoS, \( M_{\text{obs,0740}} = 2.08 \, M_\odot \), and \( \sigma_{0740} = 0.07 \, M_\odot \) is the observed uncertainty.

For PSR J0952-0607, which is much more rapidly rotating, we must account for the rotational corrections in the mass. The likelihood is:
\begin{equation}
    \mathcal{L}_{0952} \propto \exp\left(-\frac{(M_{\text{0952}}(\theta, \Omega) - M_{\text{obs,0952}})^2}{2\sigma_{0952}^2}\right),
\end{equation}
where \( M_{\text{0952}}(\theta, \Omega) \) is the mass of PSR J0952-0607, corrected for its rotational effects based on the EoS parameter \( \theta \) and spin frequency \( \Omega = 709.2 \, \text{Hz} \) with negligible error.  The correction for rotational effects follows from the relations in the rapidly rotating neutron star section. The observed mass of PSR J0952-0607 is \( M_{\text{obs,0952}} \) with uncertainty \( \sigma_{0952} \).
The rotationally corrected mass for PSR J0952-0607 is:
\begin{equation}
    M_{\text{0952}}(\theta, \Omega) = M_*(\theta) + \Delta M(\theta, \Omega),
\end{equation}
where \( M_*(\theta) \) is the mass of a non-rotating neutron star, and \( \Delta M(\theta, \Omega) \) accounts for rotational effects.

\subsection{Priors and Sampling}
We consider uniform priors for all EoS parameters, $\log (p_1/\mathrm{dyn}~ \mathrm{cm}^{-2}) \in (33.5,34.8)$, $\Gamma_1 \in (1.4,5)$, $\Gamma_2 \in (1.4,5)$, and $\Gamma_3 \in (1.4,5)$. These parameters span a range consistent with both theoretical and observational constraints on the EoS~\cite{Carney:2018sdv}.
We use the nested sampling algorithm implemented in~{\tt Pymultinest}~\citep{Buchner:2014nha} to sample the posterior distribution. This method efficiently explores the parameter space and handles high-dimensional likelihood functions, providing robust estimates of the EoS parameters.

\subsection{Scope and Limitations}
This study is focused on the impact of PSR J0952-0607’s mass and spin measurements on EoS inference. It is not intended to be a full-scale EoS inference, as we do not incorporate additional constraints from gravitational wave observations or NICER mass-radius measurements. Instead, we isolate the contribution of these two high-mass neutron stars to the EoS inference. Future studies may expand on this analysis by incorporating additional observational constraints. By limiting the scope to these two neutron stars, we can directly evaluate how their mass and spin measurements affect our understanding of the dense matter EoS, particularly at high densities. For a review of the current understanding on the EoS of NSs we refer the reader to the following articles~\cite{Raaijmakers:2019dks,Landry_2020PhRvD.101l3007L,Jiang:2019rcw,Traversi:2020aaa,Dietrich:2020efo,Miller:2021qha, Biswas:2020puz, Biswas:2021pvm, Biswas:2021yge, Biswas:2021paf, Tiwari:2023tkj, Biswas:2024hja,Char:2024kgo}.

\section{Results}
\label{sec:result}
\subsection{Constraints on Neutron Star EoS Parameters and their Properties}


We conduct a Bayesian analysis using three distinct observational constraint sets to assess whether accounting for the rapid spin of PSR J0952-0607 improves the accuracy of EoS parameter inferences. The constraint sets are:

\begin{enumerate}
    \item The mass measurement of PSR J0740+6620, \(2.08 \pm 0.07 \, M_\odot\);
    \item Combined mass measurements from PSR J0740+6620 and PSR J0952-0607, assuming PSR J0952-0607 is nonrotating with a mass of \(2.35 \pm 0.17 \, M_\odot\), to evaluate potential biases when neglecting rotation;
    \item Mass and spin constraints for PSR J0952-0607, treated as a rapidly rotating neutron star (spin of 709.2 Hz), along with PSR J0740+6620’s mass measurement. 
   \\
\end{enumerate}

\begin{figure}[h!]
\centering
\includegraphics[width=1.0\linewidth]{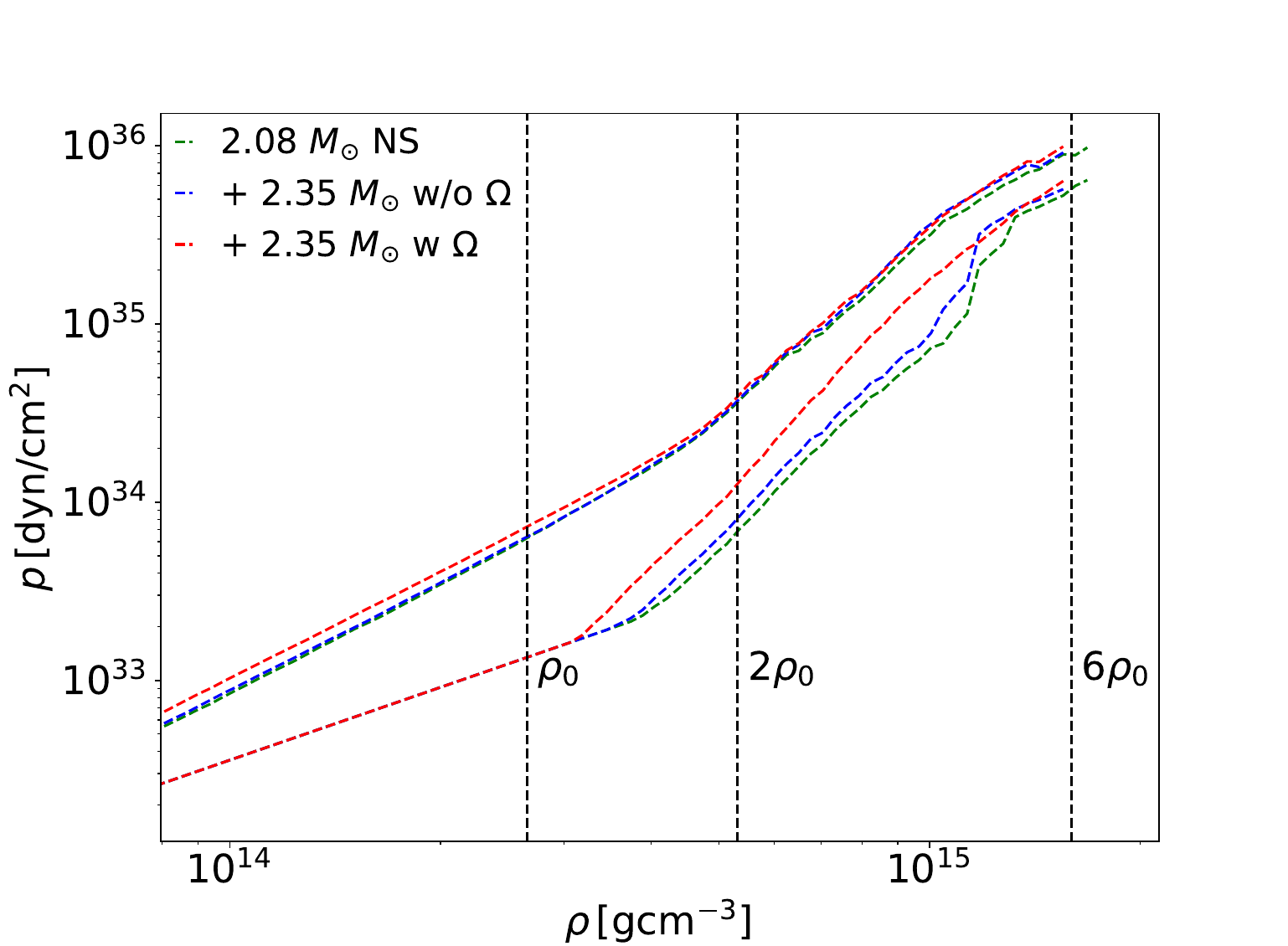}
\caption{90 \% CI of the marginalized posterior distribution of pressure vs density is shown by adding different types of constraints successively one after another, as indicated in the legend of the plot.}
\label{fig:constraint_p_d}
\end{figure}

\begin{figure}[h!]
\centering
\includegraphics[width=1.0\linewidth]{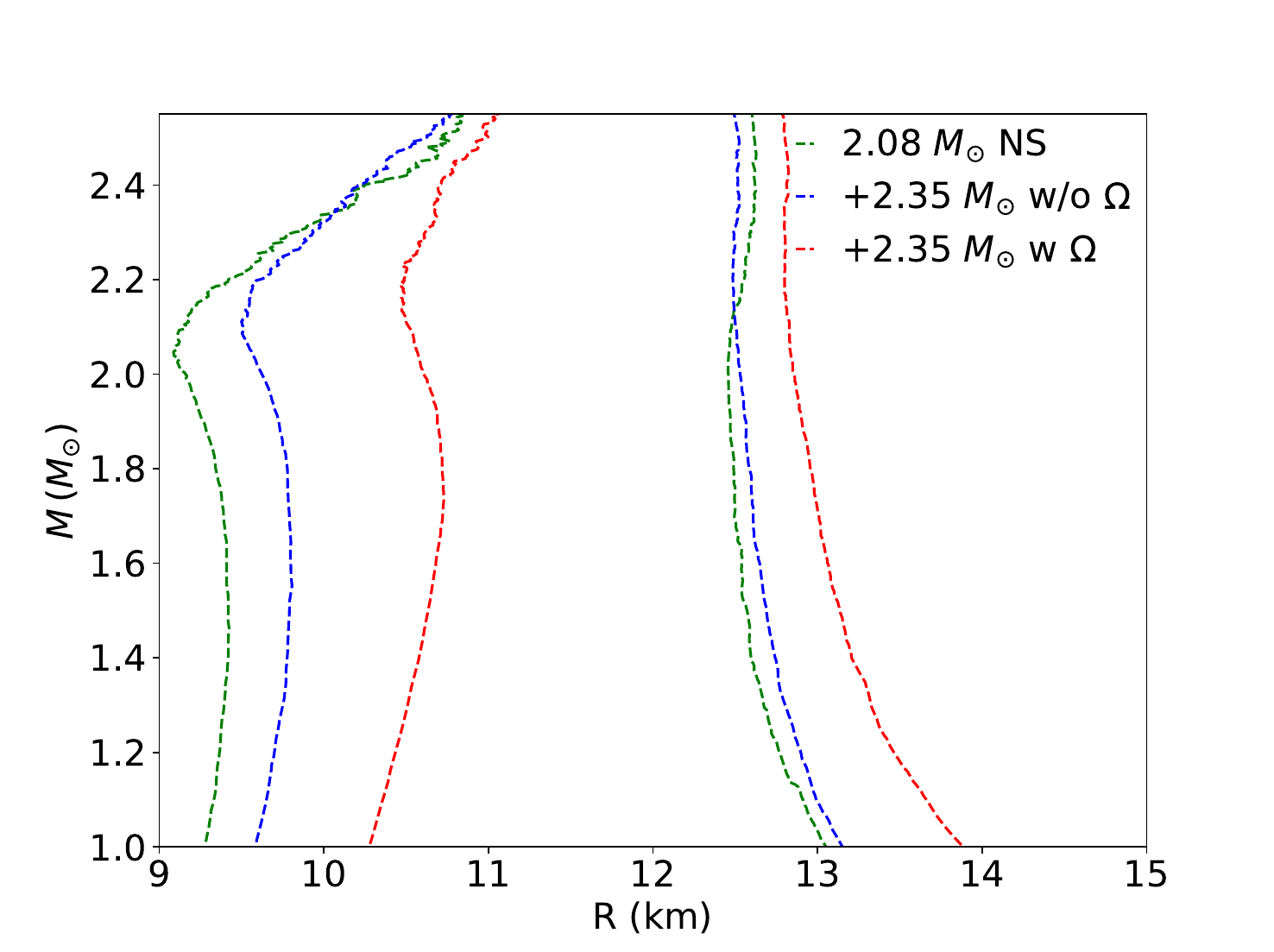}
\caption{90 \% CI of the marginalized posterior distribution of mass vs radius is shown by adding different types of constraints successively one after another, as indicated in the legend of the plot.}
\label{fig:constraint_r-m}
\end{figure}

\begin{figure}[h!]
\centering
\includegraphics[width=1.0\linewidth]{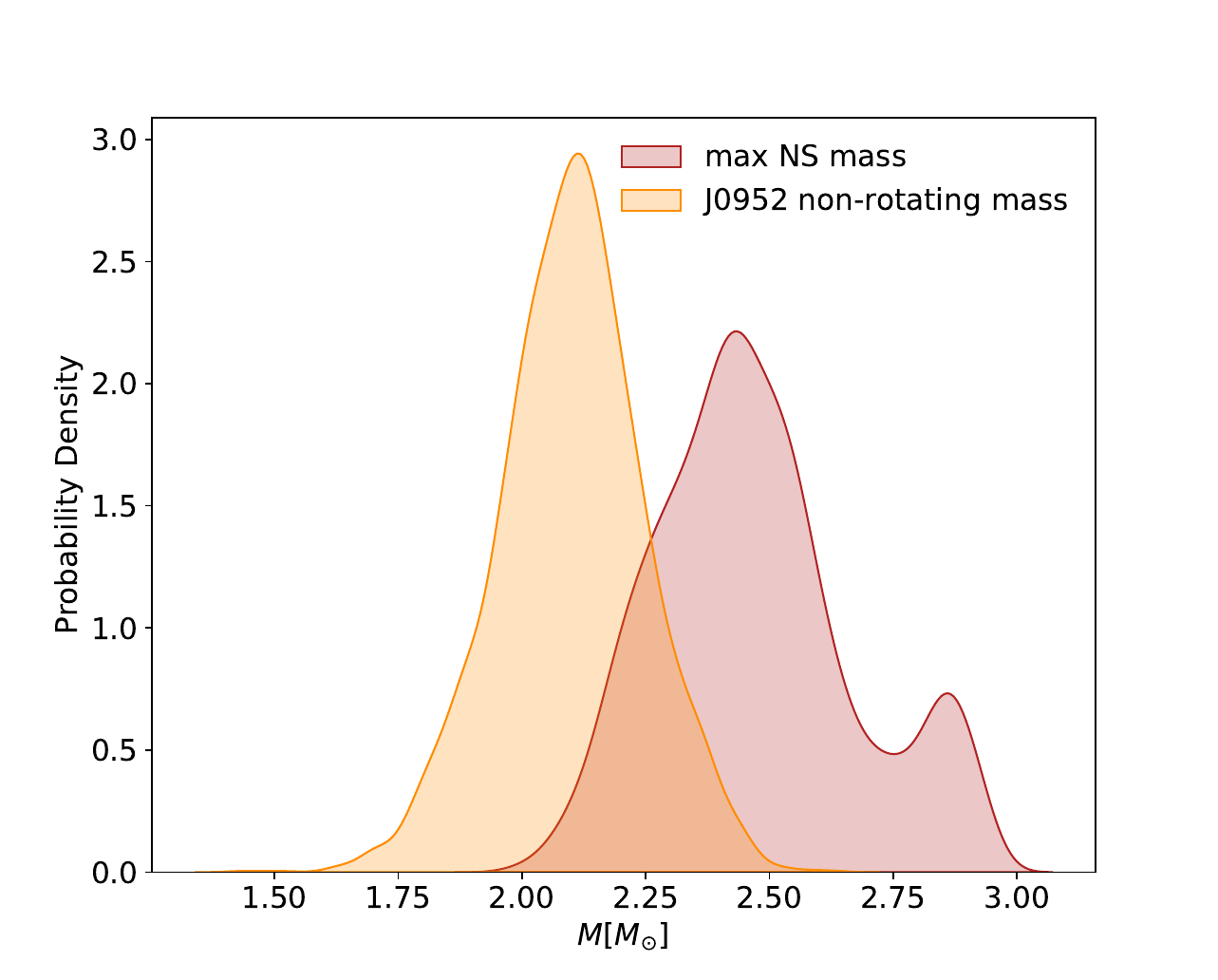}
\caption{Comparing the distributions of  TOV mass of PSR J0952-0607 and maximum mass  of nonrotating neutron stars.}
\label{fig: mass_distiburion}
\end{figure}

The Bayesian analysis results, shown in Figures~\ref{fig:constraint_p_d} and ~\ref{fig:constraint_r-m}, reveal how successive observational constraints affect the inferred neutron star equation of state (EoS), including both the pressure-density and mass-radius relations.  A few key quantities such as the radius of a $1.4 M_\odot$ and $2.08 M_\odot$ mass neutron star, and the maximum mass of nonrotating neutron stars are presented in Table \ref{tab:result-summary}:\\
\begin{table}[h]\small
\begin{center}
    \begin{tabular}{cccccc}
    \toprule
    Quantity & $2.08 M_\odot$ &$+2.35 M_\odot$ (w spin) &  $+2.35 M_\odot$ (w/o spin) \\  
    \midrule
    \\
     $R_{1.4} [\rm km]$&$10.98^{+3.19}_{-1.56}$ & $11.65^{+2.65}_{-1.08}$& $11.25^{+2.98}_{-1.47}$\\
     \\
    $R_{2.08} [\rm km]$&$10.81^{+3.35}_{-1.70}$ &$11.65^{+2.28}_{-1.10}$& $11.25^{+2.99}_{-1.73}$ \\
    \\
   $M_{\rm max} (M_{\odot})$ &$2.26^{+0.49}_{-0.21}$ &$2.45^{+0.42}_{-0.27}$&$2.40^{+0.33}_{-0.25}$\\ 
   \\
   
     \bottomrule
    \end{tabular}
    \caption{Median and $90 \%$ confidence interval (CI) of   $R_{1.4}$, $R_{2.08}$ and maximum mass $\rm\ M_{max}$ are quoted here.}
    \label{tab:result-summary}
\end{center}

\end{table}

With only PSR J0740+6620’s mass measurement, the posterior distributions for pressure and radius are broad, especially at higher energy densities and masses. For instance, the median radius for a canonical \(1.4 \, M_\odot\) NS, \( R_{1.4} \), is \(10.98^{+3.19}_{-1.56}\) km, and for a \(2.08 \, M_\odot\) NS, \( R_{2.08} \), it is \(10.81^{+3.35}_{-1.70}\) km. Similarly, the pressure posterior shows substantial uncertainty across all densities, indicating that a single mass measurement provides limited constraint on the neutron star EoS.

When we add PSR J0952-0607’s mass measurement (2.35 \( \pm \) 0.17 \( M_\odot\)) while assuming it is nonrotating, the posterior distributions narrow slightly, particularly at high densities and masses. This additional mass constraint suggests higher pressures at greater densities, reflecting the need to support more massive stars. Under this assumption, the median radius for a \(1.4 \, M_\odot\) NS, \( R_{1.4} \), shifts to \(11.25^{+2.98}_{-1.47}\) km, while \( R_{2.08} \) shifts to \(11.25^{+2.99}_{-1.73}\) km. The maximum mass also increases, with \( M_{\rm max} \) reaching \(2.40^{+0.33}_{-0.25} \, M_\odot\). However, treating PSR J0952-0607 as nonrotating may introduce a bias, as rotational effects are significant for this rapidly spinning NS and affect both mass and radius estimates, particularly at high densities.

Including both the mass and spin of PSR J0952-0607 leads to the tightest posterior distributions for the pressure-density and mass-radius relations. The additional spin constraint allows us to more accurately model the impact of rapid rotation on PSR J0952-0607’s structure, yielding narrower 90\% CIs and refining the EoS across all densities. For a \(1.4 \, M_\odot\) NS, the median radius \( R_{1.4} \) increases to \(11.65^{+2.65}_{-1.08}\) km, while for a \(2.08 \, M_\odot\) NS, \( R_{2.08} \) increases to \(11.65^{+2.28}_{-1.10}\) km. Additionally, the maximum mass \( M_{\rm max} \) increases to \(2.45^{+0.42}_{-0.27} \, M_\odot\), with reduced uncertainty across the mass and density ranges. These findings illustrate the importance of accounting for rotational effects in high-spin neutron stars like PSR J0952-0607, as failing to include spin can underestimate pressures and radii at high densities.

Finally, with the rotational constraint included, we infer the equivalent nonrotating (TOV) mass of PSR J0952-0607 to better understand its place in the NS stability landscape. The inferred value of the nonrotating mass is \(2.10^{+0.25}_{-0.24} \, M_\odot\), as shown in Fig.~\ref{fig: mass_distiburion}, where it is compared against the maximum mass distribution of NSs. This value indicates that, although PSR J0952-0607 is highly massive, it does not reach the maximum mass threshold established by the EoS (around \(2.45 \, M_\odot\)). This difference between the inferred nonrotating  mass and the maximum mass emphasizes the critical role of rapid rotation in supporting higher masses in neutron stars, as rotational effects contribute significantly to PSR J0952-0607’s observed mass. This finding illustrates that accurately accounting for rapid spin is essential for understanding the structure and stability limits of massive neutron stars.

In summary, incorporating the spin of PSR J0952-0607 provides a more precise and reliable inference of the EoS parameters, tightening constraints on the pressure-density and mass-radius relations, particularly at high densities and masses. The successive narrowing of the EoS's Confidence Intervalls (CIs)  highlights the value of adding spin information for high-spin neutron stars, which plays an essential role in the accurate modeling of NS properties and stability.

\subsection{Constraints of the r-mode instability of Neutron Stars}

\begin{figure}[h!]
\centering
\includegraphics[width=0.9\linewidth]{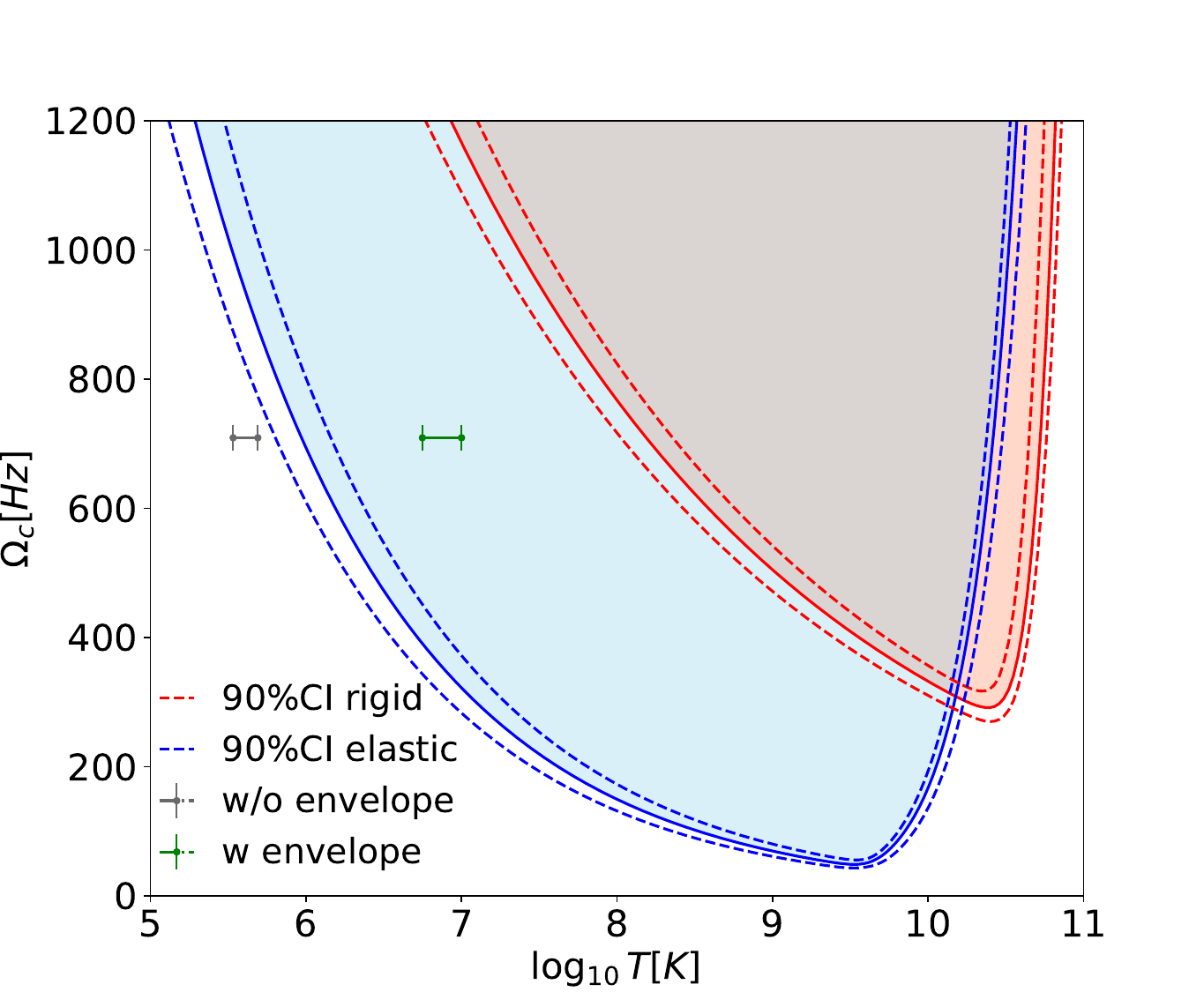}
\caption{Inferred ranges of the r-mode instability window under the assumption of NS having a rigid or elastic crust. The light blue and light red regions represent the r-mode instability ranges of a NS with an elastic crust and a rigid crust, respectively. The shaded overlapping area highlights the instability range where the NS crust is r-mode unstable, irrespective of whether it's crust is rigid or elastic. } 
\label{fig:Bayesian_omega}
\end{figure}

\begin{figure}[h!]
\centering
\includegraphics[width=0.9\linewidth]{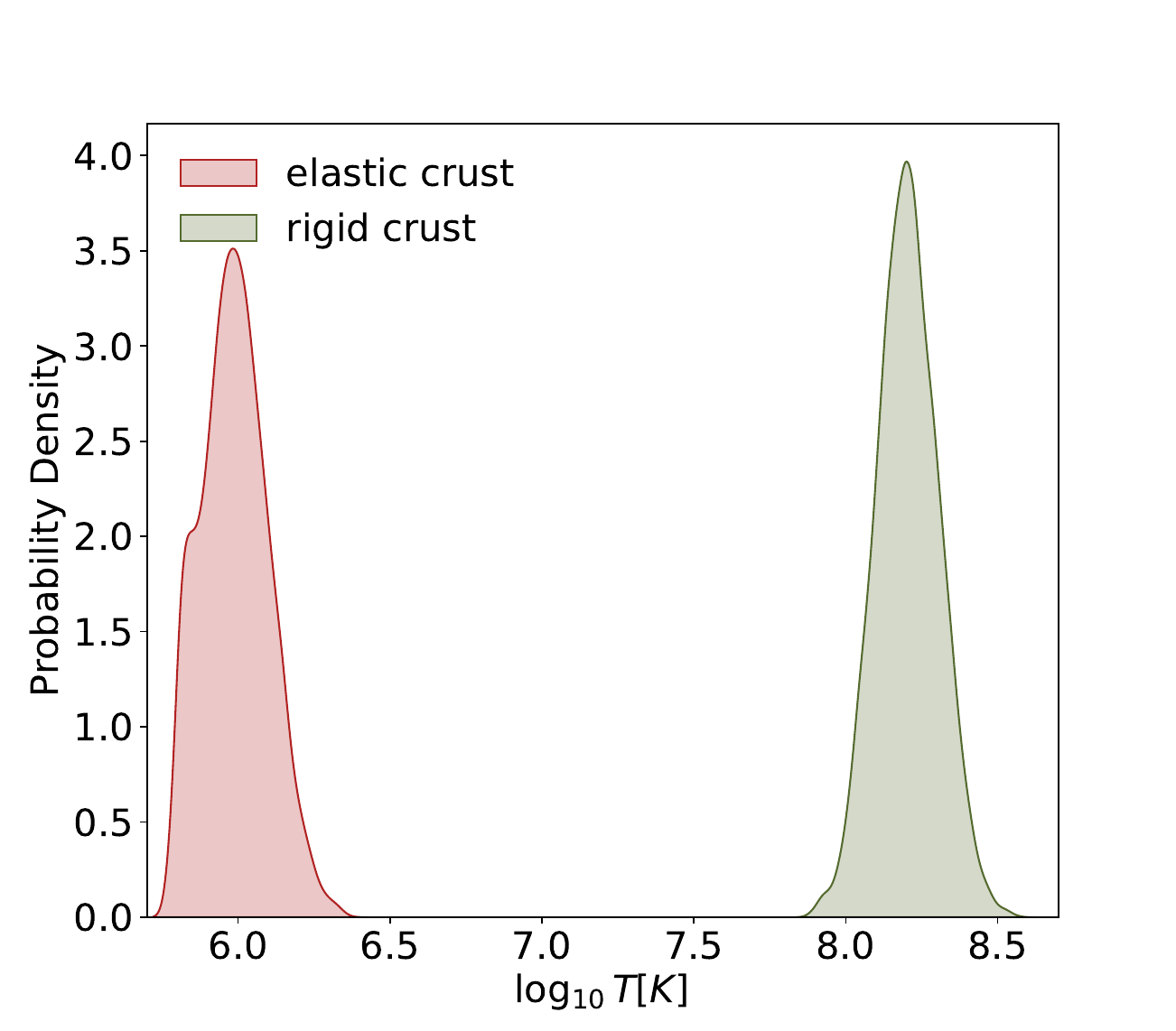}
\caption{Boundary temperature distribution for r-mode stability in both rigid and elastic crust scenarios.} 
\label{T_distribution}
\end{figure} 

Building on the combined EoS constraints derived from PSR J0740+6620 and PSR J0952-0607 (including its rapid spin), we analyze the r-mode instability window for PSR J0952-0607 under both rigid and elastic crust conditions. 
We have computed the r-mode instability window for PSR J0952-0607 for each inferred EoSs under both crust assumptions. The 90 \% credible regions of these two unstable windows are shown in Figure~\ref{fig:Bayesian_omega}. To further understand the stability behavior, we examine the boundary temperature distribution for the r-mode stability under both rigid and elastic crust conditions, shown in Figure \ref{T_distribution}. This boundary temperature represents the maximum temperature at which PSR J0952-0607 can maintain r-mode stability. Based on the combined EoS constraints, we find that the boundary temperature for the rigid crust case is significantly higher than for the elastic crust, indicating that PSR J0952-0607 would have a much larger stable region for r-modes with a rigid crust than with an elastic crust.

Next, we compare these two instability windows with the surface temperature of PSR J0952-0607 estimated by \cite{Ho:2019myl}. The surface temperature, namely effective temperature $T_e$ of a neutron star is generally lower than its core temperature $T_c$. Before the neutron star reaches thermal equilibrium, and without considering the effects of its envelope, their relationship is given by \cite{Shapiro1983Black}:
\begin{equation} 
\frac{T_e}{T_c}=10^{-2} \left(\frac{10^9K}{T_c}\right)^{\frac{5}{8}} \left(\frac{M}{M_\odot}\right)^{\frac{1}{4}} \left(\frac{R}{10 \rm km}\right)^{-\frac{1}{2}}
\label{T_e}
\end{equation}
Once thermal equilibrium is reached,  the surface temperature $T_e$ is assumed to be equal to the core temperature $T_c$. ~\citet{Ho:2019myl} find that, without envelope effects, the core temperature aligns with the surface temperature in the range $3.4 \times 10^5K<T_c<4.9 \times 10^5K$. However, the insulation effect of the envelope, a layer approximately tens to hundreds of meters thick located outside the outer crust of a neutron star, cannot be neglected. This envelope is primarily composed of either heavy elements, such as iron, or light elements, such as hydrogen. If the envelope consists mainly of heavy elements, its thermal conductivity is relatively low, resulting in a stronger insulating effect that limits the heat transfer from the core to the surface. In contrast, an envelope dominated by light elements has higher thermal conductivity, allowing more efficient heat transfer and consequently weaker insulation \cite{Ho:2019myl}.

For these two types of envelopes, the relationship between \(T_{\text{e}}\) (surface temperature  and \(T_c\) (core temperature) is given as follows \cite{Ho:2019myl}:

\begin{itemize}
    \item For an iron envelope:
    \begin{equation}
    T_c = 1.29 \times 10^8 \, g_{14}^{1/4} \, T_{\text{e},6}^{0.55} \, \text{K}
    \end{equation}
    \item For a fully accreted hydrogen envelope:
    \begin{equation}
    T_c = 0.552 \times 10^8 \, g_{14}^{1/4} \, T_{\text{e},6}^{0.7} \, \text{K}
    \end{equation}
\end{itemize}
where \(T_c\) is the core temperature,  
\(T_{\text{e}}\) is the surface effective temperature,  
\(g_{14}\) is the surface gravity, in units of $10^{14} {\rm cm/s}^2$,  
\(T_{\text{e},6}\) is the surface temperature, in units of \(10^6 \, \text{K}\). Based on these formulae, considering the insulation effect of the envelope increases the estimated core temperature range to \(5.6 \times 10^6 \, \text{K} < T_c < 1 \times 10^7 \, \text{K}\) \cite{Ho:2019myl}. Comparing this temperature range to the instability window, we find that PSR J0952-0607 is likely r-mode stable if it has a rigid crust. However, with an elastic crust and significant envelope insulation, the star would fall within the instability window, suggesting an r-mode unstable state.\\

\section{Conclusion}
\label{sec:conclusion}
This study utilizes the distinctive properties of PSR J0952-0607—its high mass of \( 2.35 \pm 0.17 \, M_{\odot} \) and rapid spin of 709.2 Hz—to refine our understanding of the neutron star equation of state and probe mechanisms that might suppress r-mode instabilities. By incorporating rotational corrections in Bayesian inferences for EoS parameters, we provide tighter constraints on pressure-density and mass-radius relations at high densities, emphasizing the importance of accounting for spin in massive, rapidly rotating neutron stars.

Our analysis of r-mode instability for PSR J0952-0607 highlights the challenges in stabilizing the star against r-mode oscillations. We examined two crust models—a rigid crust and an elastic crust—to explore their potential effects on r-mode instability. Results indicate that a rigid crust provides a larger stability window, allowing PSR J0952-0607 to remain r-mode stable at higher core temperatures than the elastic crust, which tends to make the star more susceptible to instability. 

A major caveat of our work is the assumption that the rigid crust of neutron stars provides the strongest r-mode damping among a multitude of possible damping mechanisms. While our current work favors the dominating rigid crust damping, there may be other even stronger damping mechanisms and/or model ingredients such as dissipation in the Ekman layer \cite{Bildsten:1999zn, Levin:2000vq, Glampedakis:2004ac}, mutual friction \cite{Haskell_2009}, resonant $r$-mode stabilization by superfluid modes \cite{Kantor:2020dex, Kantor:2021ipr} and enhanced bulk viscosity in hyperonic matter \cite{Nayyar:2005th, Ofengeim:2019fjy} that significantly affect the location of the r-mode instability windows. Investigating these stabilizing factors will deepen our understanding of the physical conditions required to achieve r-mode stability in massive, high-spin neutron stars, thus advancing our knowledge of dense matter under extreme astrophysical conditions. But this is left for future studies.

\section*{Acknowledgements}

We are extremely thankful to Nikolaos Stergioulas and Prasanta Char for reading this manuscript carefully and making several useful suggestions.
  BB  acknowledges the support from the Knut and Alice Wallenberg Foundation 
under grant Dnr.~KAW~2019.0112 and the Deutsche 
Forschungsgemeinschaft (DFG, German Research Foundation) under 
Germany's Excellence Strategy – EXC~2121 ``Quantum Universe'' –
390833306. SR has been supported by the Swedish Research Council (VR) under grant number 2020-05044, by the research environment grant “Gravitational Radiation and Electromagnetic Astrophysical Transients” (GREAT) funded by the Swedish Research Council (VR) under Dnr 2016-06012, by the Knut and Alice Wallenberg Foundation under grant Dnr. KAW 2019.0112, by Deutsche Forschungsgemeinschaft (DFG, German Research Foundation) under Germany’s Excellence Strategy - EXC 2121 “Quantum Universe” - 390833306 and by the European Research Council (ERC) Advanced Grant INSPIRATION under the European Union’s Horizon 2020 research and innovation programme (Grant agreement No. 101053985). 

\bibliography{mybiblio}
\end{document}